\documentclass[twocolumn,showpacs,amsmath,amssymb]{revtex4}
\usepackage{graphicx}
\usepackage[latin1]{inputenc}
\usepackage{dcolumn}
\usepackage{bm}
\usepackage{epsfig}
\usepackage{color}
\usepackage{hyperref}

\begin{document}

\preprint{APS/123-QED}

\title{Speech earthquakes: scaling and universality in human voice}

 \author{Jordi Luque$^1$, Bartolo Luque$^2$, and Lucas Lacasa$^{3,*}$}
\affiliation{$^1$Telefonica Research, Edificio Telefonica-Diagonal 00, Barcelona, Spain\\
$^2$Departamento de Matem\'atica Aplicada y Estad\'istica, EIAE, Universidad Polit\'ecnica de Madrid, Spain\\
$^3$School of Mathematical Sciences, Queen Mary University of London,
Mile End Road E14NS, London, United Kingdom}
\email{
l.lacasa@qmul.ac.uk}

\date{\today}

\begin{abstract}
Speech is a distinctive complex feature of human capabilities. In order to understand the physics underlying speech production, in this work we empirically analyse the statistics of large human speech datasets ranging several languages. We first show that during speech the energy is unevenly released and power-law distributed, reporting a universal robust Gutenberg-Richter-like law in speech. We further show that such earthquakes in speech show temporal correlations, as the interevent statistics are again power-law distributed. Since this feature takes place in the intraphoneme range, we conjecture that the responsible for this complex phenomenon is not cognitive, but it resides on the physiological speech production mechanism. Moreover, we show that these waiting time distributions are scale invariant under a renormalisation group transformation, suggesting that the process of speech generation is indeed operating close to a critical point. These results are put in contrast with current paradigms in speech processing, which point towards low dimensional deterministic chaos as the origin of nonlinear traits in speech fluctuations. As these latter fluctuations are indeed the aspects that humanize synthetic speech, these findings may have an impact in future speech synthesis technologies.
Results are robust and independent of the communication language or the number of speakers, pointing towards an universal pattern and yet another hint of complexity in human speech.
\end{abstract}

 \maketitle

The description, understanding and modelling of speech is an interdisciplinary topic of current
interest for physics \cite{Hegger}, social and cognitive sciences \cite{Hickok, Kello-2008}, data mining as well as
engineering \cite{Wilson, engineering, citeulike:11988920,Taylor:2009}.
Classical speech synthesis technologies and algorithms were firstly based on linear stochastic models and linear prediction \cite{citeulike:11988920,Taylor:2009,rabiner_schafer78}, and their underlying theory, so called source-filter theory of speech production, was initially relying on several key assumptions including uncoupled vocal tract and speech source, laminar airflow propagating linearly, periodic fold vibration, and homogeneous tract conditions \cite{rabiner_schafer78,citeulike:11988920,KentandRead}. Despite the successes of this benchmark theory, the synthetic speech generated by linear models fails to be `natural'. As a matter of fact, current speech synthesizers usually require to incorporate pieces of real speech, e.g. concatenation of smaller speech units (phoneme and diphone-based synthesis \cite{citeulike:11988920,Taylor:2009}) to improve their synthetic output.\\
With the popularisation of nonlinear dynamics, fractals, and chaos theory, the modelling paradigm slightly shifted and nonlinear speech processing emerged \cite{marshall2006advances}. Accordingly, a number of authors pointed towards low dimensional, chaotic phenomena as the underlying mechanism governing the fluctuations in speech. This paradigm shift has fostered a number of inspiring theoretical vocal-fold and tract models \cite{Steinecke} of increasing complexity,
displaying a range of nonlinear phenomena such as bifurcations and chaos \cite{Jiang} or irregular oscillations \cite{Neubauer} to cite some. Glottal airflow induced by the fluid-tissue interaction has also been modelled using Navier-Stokes equations \cite{Tao-2008}. Nevertheless, the empirical justification for low dimensional chaos is, up to a certain extent, based on preliminary evidence \cite{Kumar-1996, Maragos-1999}, and this modelling approach only theoretically justified through an analogy between turbulent states, chaos, and fractals \cite{marshall2006advances}. As a matter of fact, it has been recently acknowledged in the nonlinear dynamics community that a word of caution should be taken when empirically analysing the low dimensional nature of short, noisy and nonstationary empirical signals, as not only the computation of dynamical invariants such as fractal dimensions or Lyapunov exponents is difficult in those cases \cite{kantz2004nonlinear}, but furthermore correlated stochastic noise can be misleadingly described as to have a low dimensional attractor \cite{osborne1989finite}. 
It is therefore reasonably unclear what modelling paradigm should we follow to pinpoint the nonlinear nature of the fine grained details of speech.\\
In this work we make no a priori assumptions about the underlying adequate dynamical model, and we follow a data driven approach. We thoroughly analyse the statistics of speech waveforms in extensive real datasets \cite{DBLP:conf/lrec/Rodriguez-FuentesPVDB12} that extend all the way into the intraphoneme range ($t < 10^{-2}$ s) \cite{Crystal1988}, enabling us to dissect the purely physiological aspects that are playing a role in the production of speech, from other aspects such as cognitive effects \cite{Kello-2008, Kello-2010}. Our main results are the following: (i) energy releases in speech are power law distributed with a language-independent universal exponent, and accordingly a Gutenberg-Richter-like law \cite{kristensen} is proposed within speech.
(ii) In the intraphoneme range ($t < 10^{-2}$ s), the interevent times (silences of duration $\tau$) between energy releases are also power-law distributed, suggesting long-range correlations in the time fluctuations of the amplitude signal. (iii) Furthermore, these distributions are invariant under a time Renormalisation Group (RG) transformation, \cite{Corral-2004,Corral-2005,Corral-2009,Corral-2010}). On the basis of these results, we should conclude that the physiological mechanism of speech production self-organises close to a critical point.\\

\begin{figure}
\centering
\includegraphics[width=0.45\textwidth]{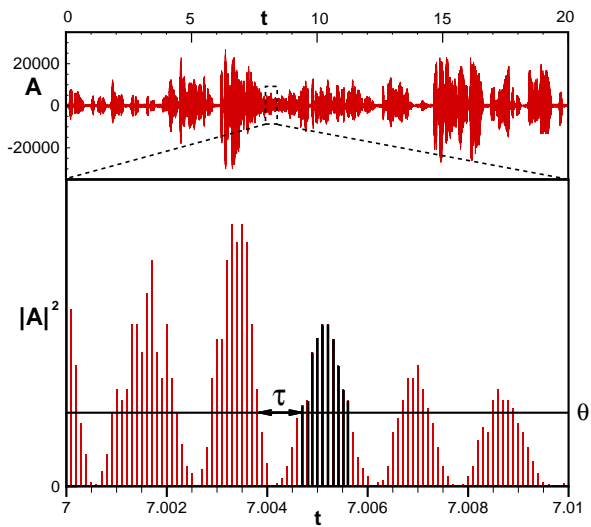}
\caption{(Top Panel) Sample speech waveform $A(t)$ of a single-speaker recording (Spanish language). (Bottom panel) Instantaneous energy per unit time $\varepsilon(t)=|A(t)|^2$ in an excerpt from the top panel. The energy threshold $\Theta$, defined as the instantaneous energy level for which a fixed percentage of the entire data remains above that level, helps us to unambiguously distinguish, for a given threshold, speech events (events for which $\varepsilon(t)>\Theta$) from silence or speech interevents of duration $\tau$. The energy released $E$ in a speech event is computed from the integration of the instantaneous energy over the duration of that event $E \sim \sum_{\text{event}} \varepsilon(t)$ (dark area in the figure denotes the energy released in a given speech event).} \label{presentacion}
 \end{figure}

\section*{RESULTS}

A TV broadcast speech database named KALAKA-2 \citep{DBLP:conf/lrec/Rodriguez-FuentesPVDB12} is employed for analysing language-dependent speech. It was originally designed for language recognition evaluation purposes and consists of wide-band TV broadcast speech recordings (roughly 4 hours per language) featuring 6 different languages: Basque, Catalan, Galician, Spanish, Portuguese and English. TV broadcast shows were recorded and sampled using $2$ bytes at $16000$ samples/second rate, taking care of including as much diversity as possible regarding speakers and speech modalities. It includes both planned and spontaneous speech throughout diverse environment conditions, such as studio or outside journalist reports but excluding telephonic channel. Therefore audio excerpts may contain voices from several speakers but only a single language. For illustrative purposes in figure \ref{presentacion} we depict a sample speech waveform amplitude $A(t)$ and its squared, semi-definite positive instantaneous energy $\varepsilon(t)=|A(t)|^2$, respectively. Without loss of generality, dropping the irrelevant constants, $\varepsilon(t)$ has units of energy per time. Then, a threshold $\Theta$ is defined as the instantaneous energy level for which a fixed percentage of data is larger than the threshold. For instance, $\Theta=70\%$ is the threshold for which $30\%$ of the data fall under this energy level (this allows to compare data across different empirical signals \cite{Danon}). $\Theta$ not only works as a threshold of `zero energy' that filters out background (environmental) noise, but help us to unambiguously distinguish a speech event, defined as a sequence of \textit{consecutive} measurements $\varepsilon(t)>\Theta$, from a silence event, whose duration is $\tau$ (see figure \ref{presentacion}). Accordingly, speech can now be seen as a dynamical process of energy releases or `speech earthquakes' separated by silence events, with in principle different statistical properties given different thresholds $\Theta$. In what follows we address all these properties.\\

\begin{figure}[t]
\centering
\includegraphics[width=0.4\textwidth]{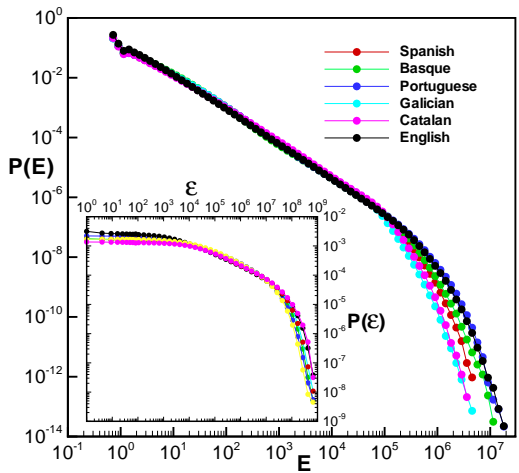}
\caption{Log-log plot of the integrated energy distribution $P(E)$ released in the statistics of all languages considered in this study, after a logarithmic binning. The figure shows a power law $P(E)\sim E^{-\alpha}$, that holds for five decades, truncated for large (rare) events by a finite-size cut-off. (Inset panel) Log-log plot of the associated instantaneous energy histogram $P(\varepsilon)$ of each language, which we include for completeness. We note that this non scaling shape of instantaneous energy is similar to the one found for precipitation rates \cite{Peters-2010}.} \label{energy}
 \end{figure}

\noindent \textbf{Energy release: a Gutenberg-Richter-like scaling law in speech}\\
The energy of a speech event is computed from the integration of the instantaneous energy over the duration of that event $E=\int_{\text{event}}\varepsilon(t) dt \approx \sum_{\text{event}}\varepsilon(t) \Delta t$ where $\Delta t=1/16000 s$ is the inverse of the sampling frequency (and therefore $E$ has arbitrary units of energy). In order to get rid of environmental noise, we set a fixed threshold $\Theta=80\%$ and for each language, we compute its histogram $P(E)$. In figure \ref{energy} we draw, in log-log scales, this histogram for all languages considered (note that a logarithmic binning was used to smooth out the data). We find out a robust power-law scaling $P(E)\sim E^{-\alpha}$ over five decades saturated by the standard finite-size cutoff, where the fitted exponents are all consistent with a language-independent universal behaviour: $\alpha=1.13\pm0.06$ for Spanish, $\alpha=1.16\pm0.05$ for Basque, $\alpha=1.16\pm0.07$ for Portuguese, $\alpha=1.13\pm0.05$ for Galician, $\alpha=1.10\pm0.05$ for Catalan, and $\alpha=1.15\pm0.02$ for English, all having a correlation coefficient $r^2>0.999$
(for completeness, in the inset of figure \ref{energy} we also depict the binned histogram of instantaneous energy $P(\varepsilon)$ for all languages).

As long as the magnitude in seismicity \cite{kristensen} is related to the logarithm of the energy release, under this definition $P(E)$ can be identified as a new 'Gutenberg-Richter-like law' in speech. This may be seen as related to other scaling laws in cognitive sciences \cite{Kello-2010}, although at this stage it is still unclear what particular contributions come from both the mechanical (vocal folds and resonating cavity) and the cognitive systems.\\

\begin{figure*}
\centering
\includegraphics[width=0.42\textwidth]{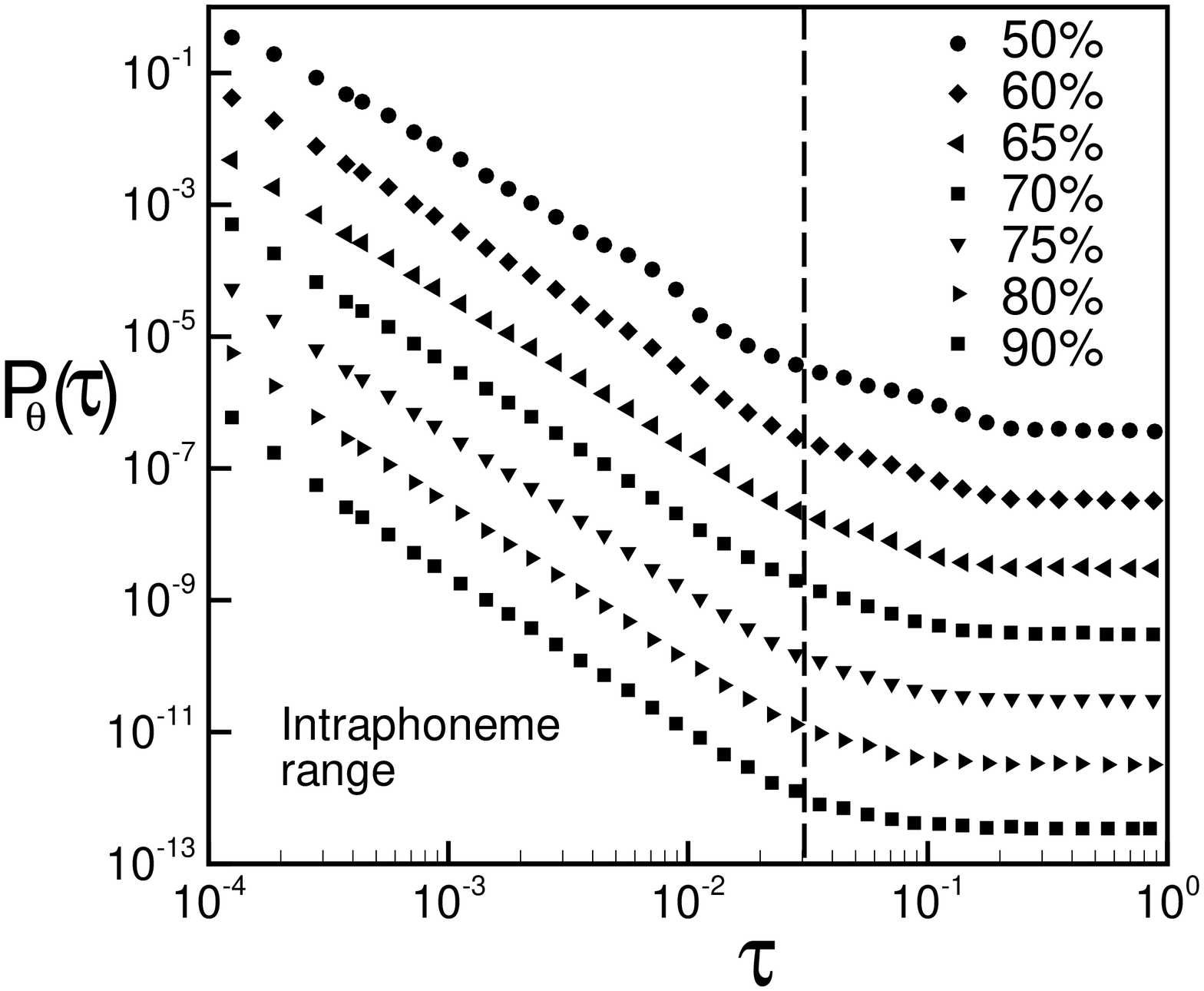}
\includegraphics[width=0.45\textwidth]{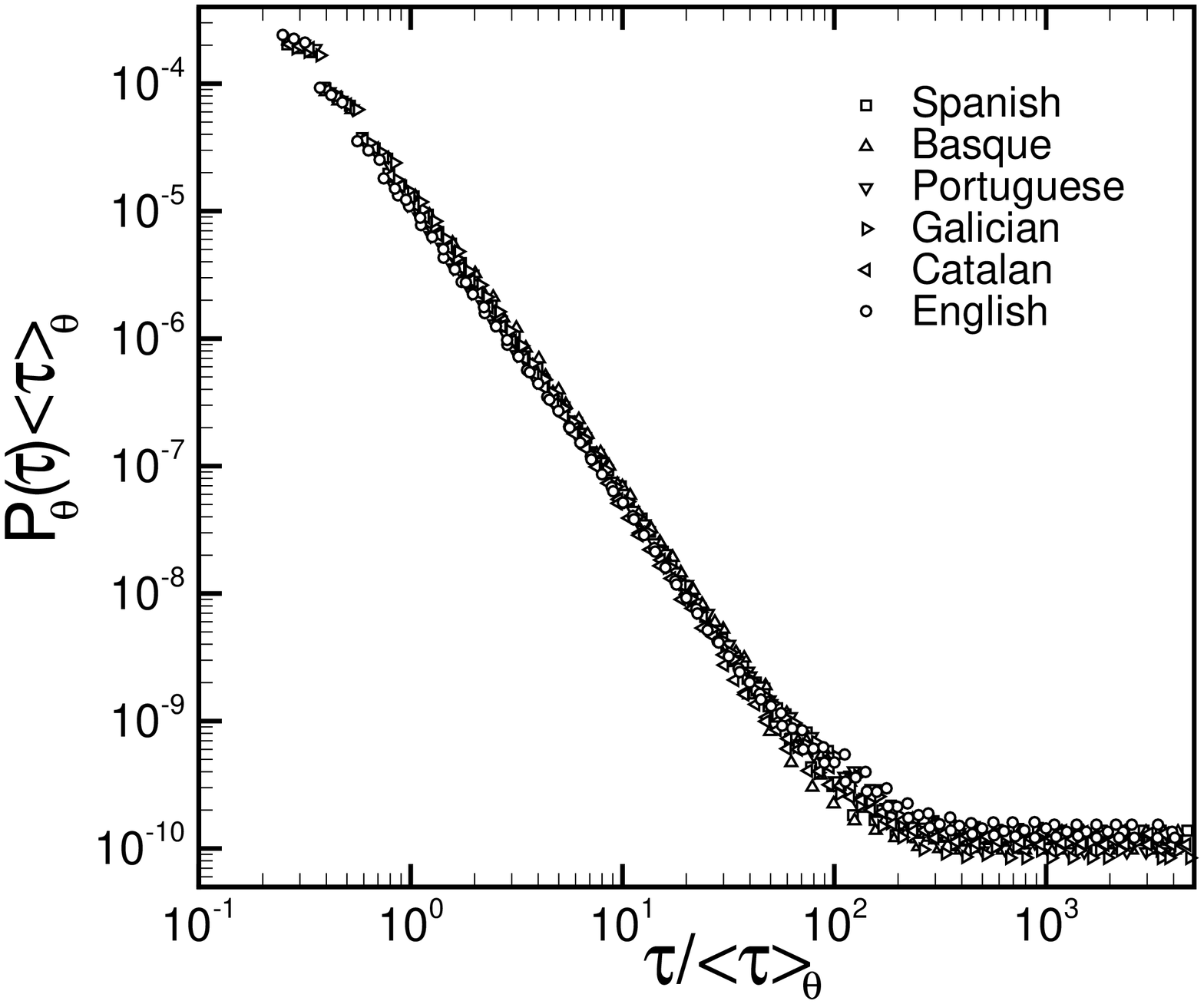}
\caption{(a) Log-log plot of the speech interevent distribution $P_{\Theta}(\tau)$ computed for the English database, for different energy thresholds $\Theta$ (there is a vertical translation performed for illustrative purposes, and data has been logarithmically binned). In the intraphoneme range (timescales $t<3\cdot10^{-2}s$), we find power law distributed interevent times fairly independently from the threshold, what suggests the onset of long-range correlations in this regime.
 (b) Rescaled interevent distributions $P(\tau)$ according to the decimation method (see the text), for all languages considered in this study and different thresholds. In the intraphoneme range, interevents collapse in a universal function and are invariant under the RG transformation, suggesting that the underlying physiological mechanism of speech synthesis is critical.} \label{interevent}
 \end{figure*}

\noindent \textbf{Scaling and universality of waiting time distributions.}\\
In a second part, we study on the temporal orchestration of fluctuations, that is, the arrangement of silences or speech interevents of duration $\tau$. We will pay a special attention to the intraphoneme range (timescales $t<3\cdot 10^{-2}$ s \cite{Crystal1988,citeulike:11988920}), where we assume no cognitive effects are present, in order to focus on the physiological aspects of speech. 
At this point we introduce a renormalisation group (RG) transformation to explore the origin of temporal correlations. This technique originates in the statistical physics community and has been previously used in the context of earthquakes \cite{Danon, Corral-2004, Corral-2005} and tropical-cyclone statistics \cite{Corral-2010}. The first part of the transformation consists of a decimation: we raise the threshold $\Theta$. This in general leads to different interevent distributions $P_{\Theta}(\tau)$. The second part of the transformation consists in a scale transformation in time, such that renormalized systems become comparable: $\tau \rightarrow   \tau /\langle\tau\rangle_\Theta$, $P(\tau)  \rightarrow  P_\Theta(\tau)\langle\tau\rangle_\Theta$, where $\langle\tau\rangle_\Theta$ is the mean interevent time of the system for a particular $\Theta$. Invariant distributions under this RG transformation collapse into a threshold-independent universal curve: an adimensional waiting time distribution. While the complete fixed point structure of this RG is not well understood yet, recent advances \citep{Corral-2005, Corral-2009} rigorously found that stable (attractive) fixed points include the exponential distribution and a somewhat exotic double power-law distribution, which are attractors for both memoryless and short-range correlated stochastic point processes under the RG flow. Invariant distributions other than the previous fixed points are likely to be unstable solutions of the RG flow, therefore encompassing criticality in the underlying dynamics \citep{Corral-2009}.\\
In the left panel of figure \ref{interevent} we plot in log-log, for different thresholds, the interevent histogram $P_{\Theta}(\tau)$ associated to the English language. In the intraphoneme range, interevents are power-law distributed in every case $P_{\Theta}(\tau)\sim \tau^{-\beta}$. In the right panel of the same figure we plot the rescaled histograms $P(\tau)$ ($\Theta = 60\%, 65\%, 70\%$ for every language, yielding a total of 18 curves), collapsing under a single curve. Note that the collapse is quite good for those timescales that belong to the intraphoneme range, where only physiological mechanisms are in place, and such collapse is lost for larger timescales. This suggests that for every language, the statistics are invariant under this RG transformation, and the pattern is robust across languages. A more careful statistical analysis is required here as the range of the power law, restricted to the intraphoneme regime, is smaller. Accordingly, exponent $\beta$ is estimated now following Clauset et al.'s method \cite{Clauset} 
which employs Maximum Likelihood Estimation (MLE) of a power law model, where goodness-of-fit test and confidence interval are based on Kolmogorov-Smirnov (KS) tests (comparing the actual distribution with 100 synthetic power law models whose exponent is the one found in the MLE to obtain p-values). This method yields a fairly universal exponent $\beta=2.06$ within the $95\%$ confidence interval $[1.84, 2.20]$, KS p-value of $0.99$ and the statistical support for the power-law hypothesis given by a p-value of $0.95$.

\section*{DISCUSSION}

It is well known that the dynamics of speech generation are complex, nonlinear, and certainly poorly explained by the benchmark source-filter theory. The fact that a Gutenberg-Richter law for the energy release probability distribution during speech emerges
opens the possibility of understanding speech production in terms of crackling noise, a highly nonlinear behaviour that was first described in condensed matter physics \cite{sethna} and latter found in a variety of natural hazards systems including earthquakes \cite{Danon}, rain \citep{Peters-2010} or solar flares \cite{solarflares1} to cite some \cite{Corral-2009}.  The underlying theory might then describe energy releases as the resonating response function of the system under airflow perturbation (the so called susceptibility in the statistical physics jargon), and the fact that it is a power law distributed quantity is the first evidence of criticality in these systems.\\
The fact that the waiting time distributions are different from the well known stable exponential laws while being invariant under the RG transformation confirms that the system is indeed operating close to a critical point. 
Note that the currently established low dimensional chaotic hypothesis suffers from being a process with short-range correlations and therefore chaotic speech interevents should typically renormalise into the exponential law. The most plausible conclusion of this work is that the physiological process of speech production evidences long-range correlations and criticality. This argument can be put in the same grounds as equilibrium critical phenomena, where only long-range correlations allow the system to escape from the basin of attraction of the trivial RG (high or low temperature) fixed points. As long as the critical solution is an \textit{unstable} fixed point of the RG flow, the fact that the underlying dynamics seems to be poised near this point is somewhat remarkable. If we assume that the dynamics of speech generation at the level of the vocal folds are influenced by the properties of the glottal airflow, then our results would suggest that a similar critical behaviour might take place in other physiological processes involving such airflow, something that has been found empirically in the mechanism of lung inflation \cite{lung}.\\ 
Note also that whereas a simple stochastic processes such as a Brownian motion cannot explain these results, under the more general paradigm of fractional Brownian motion with first return distribution $P(T)\sim T^{H-2}$, our findings would be consistent with so called  $1/f$ noise (with Hurst exponent $H=0$), on agreement with previous evidence \cite{Kello-2008}, this latter being a trait of long range correlations. This gives further credit to our results, as $1/f$ noise is usually found along with criticality \cite{kristensen}.\\
Although we acknowledge that weakly chaotic systems such as intermittent ones can operate close to criticality, our results suggest that the paradigm of self-organised criticality (SOC) \cite{kristensen} - out of equilibrium dissipative systems that self-organise towards a critical state- may be a more adequate modelling scenario than standard low dimensional chaos. If this was to be the case, threshold dynamics \cite{kristensen} would appear as the essential ingredient that encompasses the nonlinear properties in speech. These new approaches could further contribute to the development of both (i) microscopic (generative) models of speech production, and (ii) alternative methods of speech synthesis that would profit from the self-similar properties of speech at the intraphoneme range to make refined speech interpolation without needs to incorporate in the synthesis model pieces of real speech.\\
Furthermore, the combined fact that (i) the results are robust for different human languages and that (ii) the timescales involved in this analysis require that the process is purely physiological, lead us to conclude that this mechanism is a universal trait of human beings. It is an open question if the onset of SOC in this system is the result of an evolutionary process \cite{kristensen}, where human speech waveforms would have evolved to be independent of speaker and receiver distance and perception thresholds. In this context, further work should be done to investigate whether if similar patterns originate in the communication of other species, and up to which extent other variables, such as ageing, may play a role.\\

\noindent \textbf{Acknowledgments. }
The authors would like to thank Luis Javier Rodr\'{\i}guez-Fuentes and Mikel Pe{\~n}agarikano for recording and hand-labeling the speech corpus.


\end{document}